\documentclass[11pt]{article}
\usepackage{blois,epsfig}
\usepackage[latin1]{inputenc}
\usepackage[OT1]{fontenc}
\usepackage{units}
\usepackage{color}
\usepackage{pstricks,multido}

\def\Journal#1#2#3#4{{#1} {\bf #2}, #3 (#4)}

\def\NIMA{{\em Nucl. Instrum. Methods} A}

\def\PRL{\em Phys. Rev. Lett.}

\def\Proc#1#2{{\em #1} (#2)}
\def\be{\begin{equation}}
\def\ee{\end{equation}}
\def\bea{\begin{eqnarray}}
\def\eea{\end{eqnarray}}

\begin{document}
\vspace*{4cm}
\title{ THE ENERGY SPECTRUM OF ULTRA HIGH ENERGY COSMIC RAYS}

\author{ Ioana~C.~Mari\c{s} for the  Pierre Auger Collaboration}

\address{Institut f\"ur Experimentelle Kernphysik, Universit\"at
  Karlsruhe (TH), Postfach 6980, Germany}

\maketitle\abstracts{
The construction of the southern site of the Pierre Auger Observatory
is almost completed. Three independent  measurements  of the flux of the
cosmic rays with energies larger than \unit[$10^{18}$]{eV} have
been  performed during the construction phase. The surface detector
data collected until August 2007 have been used to establish  a flux
suppression at the highest energies with a
6$\sigma$ significance. The 
observations of cosmic rays by the  fluorescence detector allowed the
extension of the energy spectrum to lower energies, where the
efficiency of the surface detector is less than 100\% and a change in
the spectral index  is expected. 
}

\section{Introduction}

Cosmic rays are particles that travel through the galactic and intergalactic
space, arriving on Earth in a broad energy range up to \unit[100]{EeV}.  
Their flux drops steeply with energy, from  a few particles per
second per  m$^2$ up to one particle per km$^2$ per century at
the highest energies.
 The shape of the flux depends
on the evolution of the sources, the mechanisms that accelerate particles
up to highest energies and the energy losses during the 
propagation of the particles from the sources to earth. 
The transition from galactic to extragalactic source is 
expected to be in the \unit[0.1-5]{EeV}  energy
range.~\cite{bib:bere,bib:allard}
% To distinguish between the
%acceleration models an accurate measurement of the shape of the energy
%spectrum as well as a composition measurement are required. 

In the highest energy range, above \unit[60]{EeV}, a flux suppression
is expected due 
to the  Greisen-Zatsepin-Kuzmin effect (GZK)~\cite{bib:Greisen,bib:Zatsepin}
and/or due to the maximum energy that a cosmic ray accelerator can reach. The
GZK suppression is a propagation effect: protons 
interact with the cosmic microwave background (CMB) radiation
losing about 15\% of their energy at each encounter with these CMB
photons. Heavier elements are dissociated through photo-disintegration.

The previous experiments AGASA~\cite{bib:AGASA} and
HiRes~\cite{bib:HiRes} have given  
contradictory results regarding the ultrahigh energy 
end of the energy spectrum where
only indirect observations of cosmic rays are possible by the detection
of extensive air showers. The previous measurements of the comic rays
flux were dominated by statistical or systematic uncertainties. 
In order to decrease the uncertainties the Pierre Auger Observatory
was built as a hybrid detector, combining the two techniques employed
by the forerunner experiments: a surface detector array  and a fluorescence
detectors. 
Due to the hybrid technique the nearly calorimetric estimation of
the energy of the primary  particle as obtained from the fluorescence
technique can be transfered to the large number  of events  recorded by
the surface detector.  Recently the flux suppression  has
been seen both by the Pierre Auger  Observatory~\cite{bib:ICRC07Roth}
and by the  HiRes collaboration~\cite{bib:HiRes}.

The observatory is  described in the second
section. The surface detector data are used to deduce the energy
spectrum above \unit[3]{EeV} where the trigger efficiency is 100\%,  as
is described in the third section. Another complementary data set is
delivered by the fluorescence detector itself. It can be used to extend the
energy range down to \unit[1]{EeV}.  
This measurement is described below in the last
section together with the method of combining it with the surface detector flux
estimate. 

\section{The Pierre Auger Observatory}

After entering the atmosphere, cosmic rays interact with nuclei in the air and
start creating extensive air showers.  The charged particles that  reach the
ground are detected with the surface detector (SD), their lateral 
spread from the air shower axis at primary energies above \unit[$10^{18}$]{eV}
is of the order of a few kilometers.
On the way through the atmosphere
charged particles excite nitrogen molecules, which afterwards emit
fluorescence light in the ultra-violet band. The amount of light is
proportional to the energy deposited by the air shower in the atmosphere.

The Pierre Auger Observatory, located in the province of
Mendoza (Argentina),  is utilized to measure the properties of extensive air
showers  by observing their longitudinal development in the atmosphere as well
as their lateral spread at ground level. 
The Observatory consists of more than 1600 water-Cherenkov detectors,
filled with  
12 tonnes of water each and equipped with three  photomultipliers to detect
secondary photons and charged particles. 
The tanks  are spread  over about 3000 km$^2$ on a triangular
grid of 1.5 km spacing. The atmosphere above the array is viewed
by four fluorescence detectors (FD), each housing six telescopes, located on the
border of the area.   
The field of view of each telescope is $30^\circ$ in azimuth, and $1.5-
30^\circ$ in elevation. Light is focused with a spherical mirror
of \unit[11]{m$^2$} effective area on a camera of 440 hexagonal pixels. Each
pixel is a photomultiplier tube with
%%%%%%%%%% MU: really 18 cm2 --> check!!!
 18 cm$^2$ detection area. 
More details on
detector setup and calibration can be found
in~\cite{bib:AugerNIM04,bib:bertouCalibration}.  
An extension of the Observatory has been started with
AMIGA~\cite{bib:AMIGA}, a denser 
array of tanks equipped with muon counters which will lower the trigger
threshold energy for the SD, and 
HEAT~\cite{bib:HEAT}, three telescopes that will increase the field of view
of FD up to 60$^\circ$. 
The counterpart of the Southern side
%,a much larger experiment
is in the planning phase in the Northern hemisphere, in Lamar, Colorado,
which  will provide large statistics above \unit[50]{EeV}.

\begin{figure}
\begin{center}
\psfig{figure=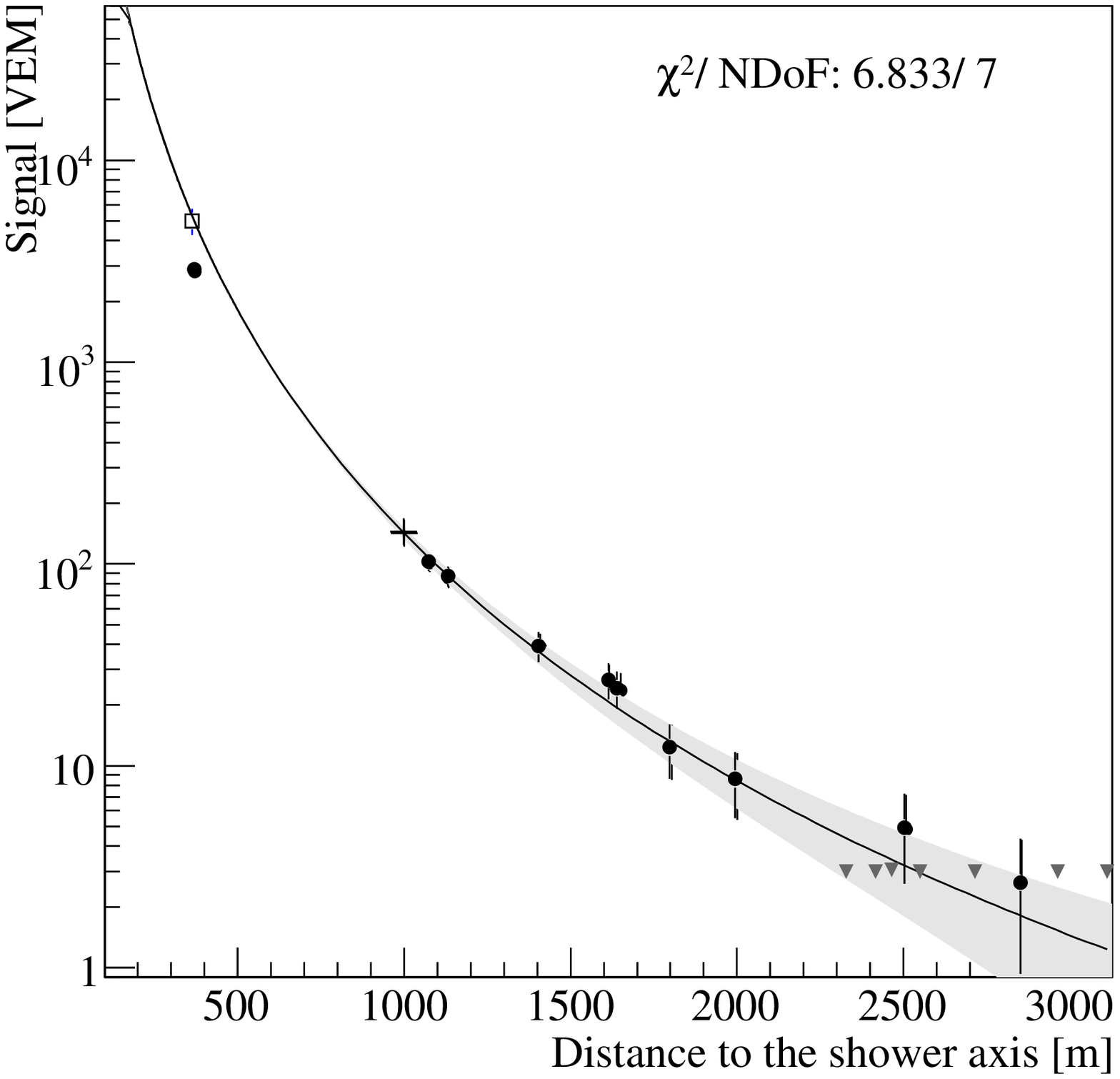,width=0.48\textwidth}
\psfig{figure=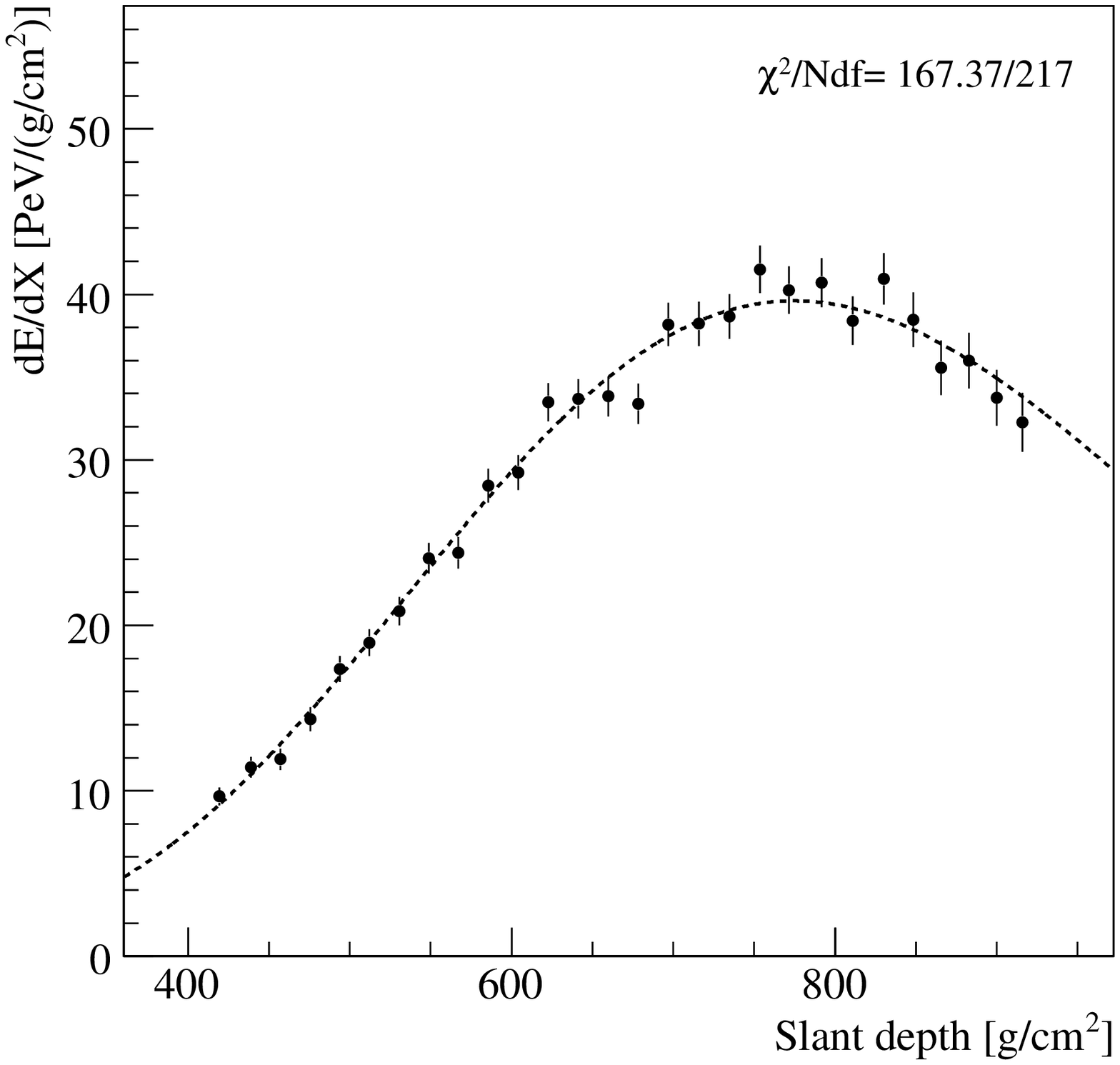,width=0.48\textwidth}
\caption{A typical golden hybrid event reconstruction, with an energy of
  \protect{\unit[30]{EeV}} and an incoming direction of $27^\circ$
  (left hand side) Lateral distribution. Filled circles represent
  acquired signals,
  triangles are functioning stations without signal used with
  Poisson  probabilities in      the maximum likelihood fit. S(1000)
  is marked with a cross.   (right     hand side) Longitudinal
  profile: energy deposit in the atmosphere  as a     function of  the
  slant depth. \hfill 
\label{fig:reco}}
\end{center}
\end{figure}

An example of a  reconstruction of the same air shower with the SD and
FD is shown in Fig.~\ref{fig:reco}. The signals recorded in the tanks
are converted  in terms of vertical equivalent muons (VEM). One VEM
represents the average of the signals  produced in the 3~PMTs  by a
vertical muon that passes centrally through the SD detector. The air
shower axis  is  obtained from the arrival 
time of the first particles in each detector station. The impact point on ground and
the lateral distribution of signals are obtained in a global maximum
likelihood minimization which  accounts for the station trigger
threshold and the overflow of  the FADCs counts in the stations very close to
the shower axis. The effect of the fluctuation of the lateral
distribution function is minimized at \unit[1000]{m}.  This optimal distance
is influenced by the array spacing. The signal at this specific
distance, S(1000), is used as energy estimator~\cite{bib:AveICRC}. 

About one in ten  air showers that reach the
surface detector are  also observed with the fluorescence detector
(the fluorescence detector operates  only in moonless 
clear nights).
The longitudinal profile of the air shower, i.e. the energy deposit as a
function of traversed matter in the atmosphere is obtained taking into
account the fluorescence  and Cherenkov light contributions and the light
scattering and attenuation~\cite{bib:ICRC07UngerFDReco}. 
Due to the limited field of view, the entire
longitudinal profile usually is no recorded, so
a fit with a Gaisser-Hillas function is employed to obtain the full profile.  
The energy of the cosmic ray is the integral over this 
function with a correction of $(10\pm 5)\%$ for the energy carried away by
the neutrinos and muons to which the FD is not
sensitive.~\cite{bib:invisible,bib:invisible2}  
The energy is proportional to the  absolute fluorescence yield in air which at
\unit[293]{K} and \unit[1013]{hPa} (\unit[337]{nm} band) is
\unit[$5.05\pm0.71$]{photons/MeV} of energy
deposited~\cite{bib:nagano}. The fluorescence yield pressure and wavelength
dependency are accounted  for~\cite{bib:airfly}.   
By using one triggered tank in the geometry
reconstruction the accuracy is improved with respect to monocular
data (i.e data recorded only by the fluorescence detector).

\section{Energy spectrum from surface detector data}

The lateral distribution of signals is a robust measurement, the only quality
criteria required for the surface detector data is that the station that
recorded the highest signal is surrounded by 6 active stations. This condition
rejects events that might be affected by the array borders.

\begin{figure}
\begin{center}
  \begin{minipage}{0.49\textwidth}
    \psfig{figure=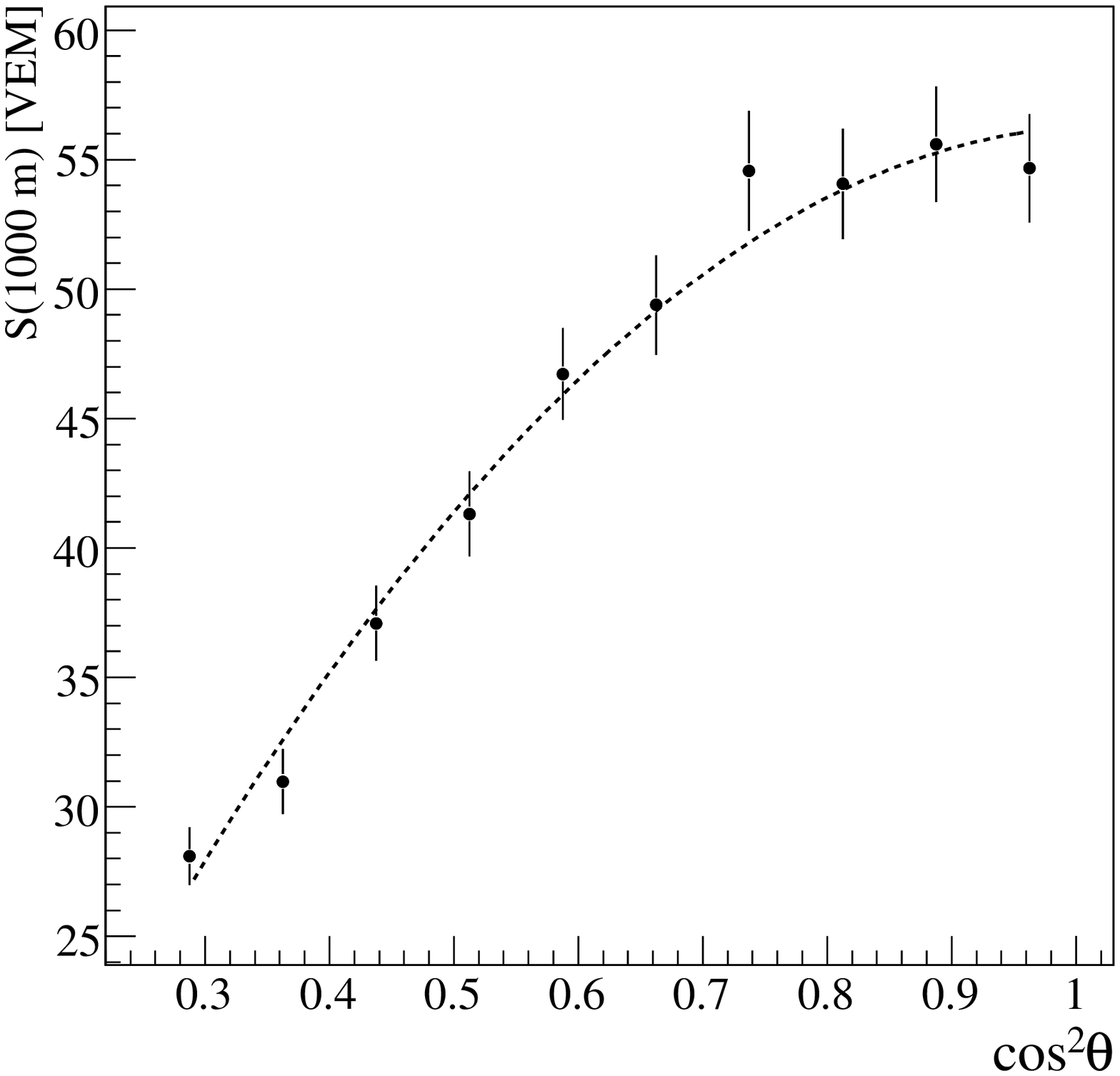,width=\textwidth}
  \end{minipage}
  \begin{minipage}{0.49\textwidth}
    \vspace*{0.5cm}
    \psfig{figure=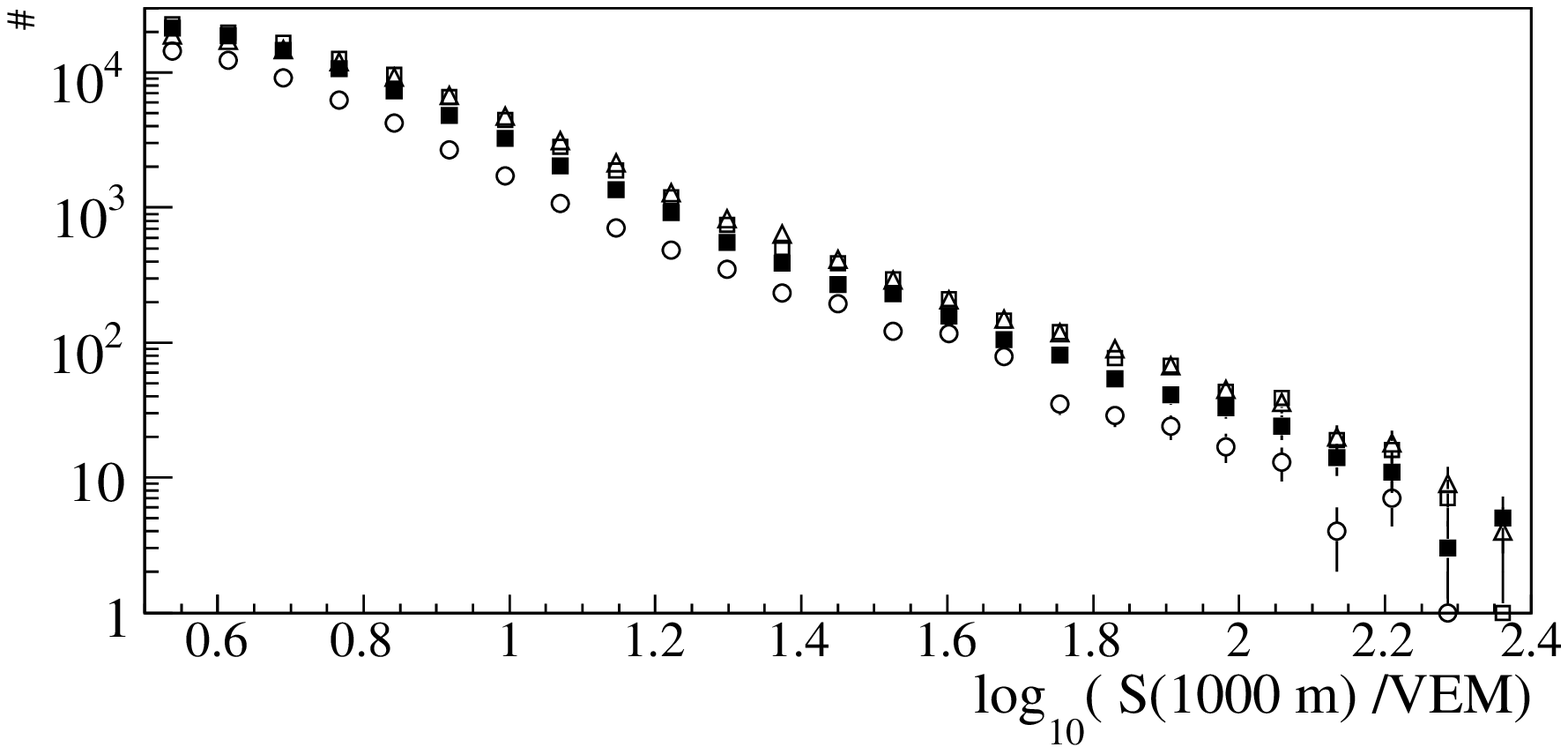,width=\textwidth}
    \psfig{figure=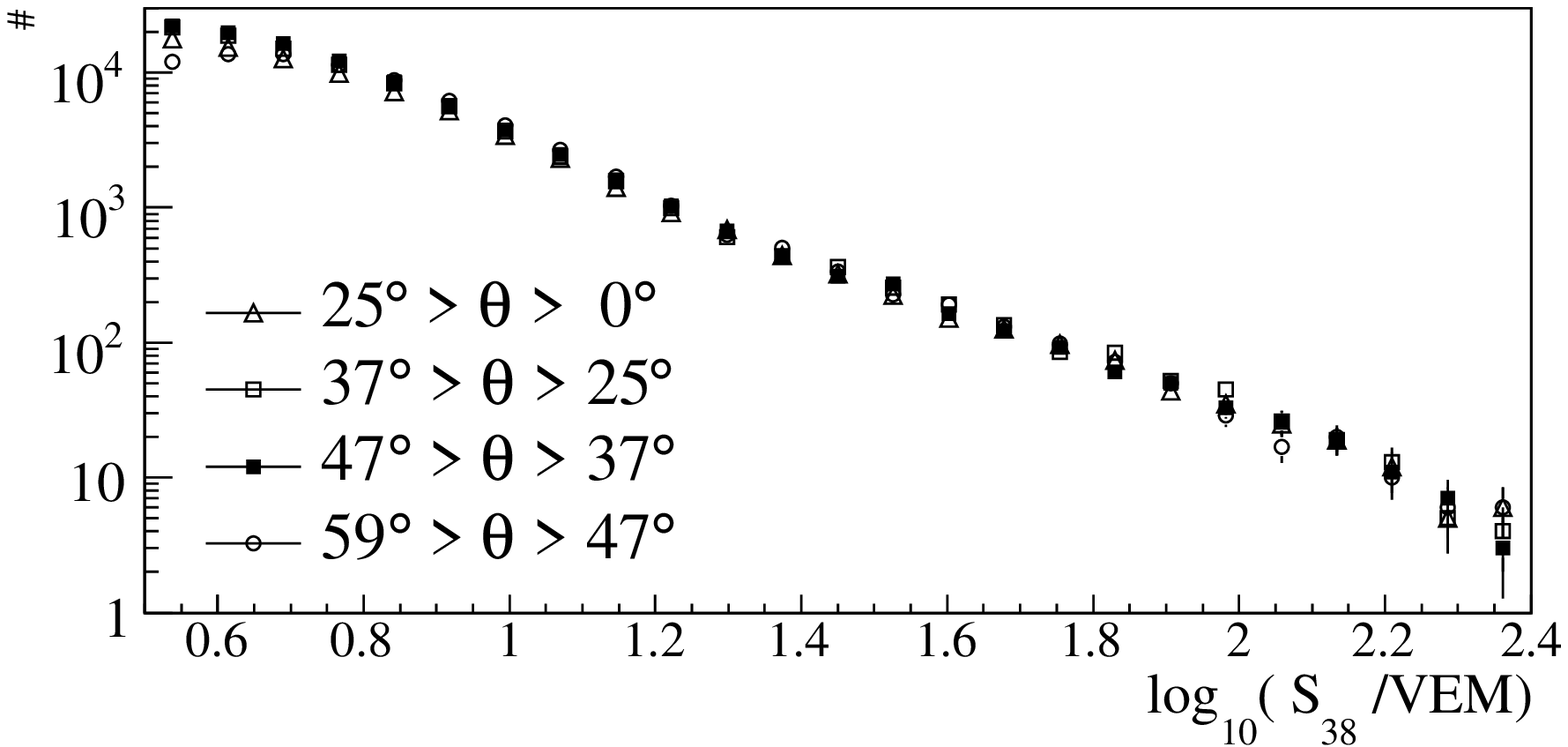,width=\textwidth}
  \end{minipage}
\end{center}
  \caption{(left hand side) S(1000) attenuation in the atmosphere. The line
    represents an   empirical fit that is used for the conversion to
    $S_{38}$.~\protect\cite{bib:ICRC07Roth} (right hand side) The number of
    events as a function of S(1000) (upper panel) and as a function of
    $S_{38}$(lower panel) for equal $\cos^2(\theta)$ intervals.\hfill   
\label{fig:cic} }

\end{figure}

The reconstructed signal at \unit[1000]{m} from the shower axis on the ground
level, S(1000), is  a good estimator for the energy of the cosmic ray.
Due to the attenuation in the atmosphere,  S(1000) depends on the  zenith
angle: an air shower developing vertically
produces a smaller signal than an inclined  shower produced by a cosmic ray
with the same energy.  
The constant intensity method~\cite{bib:ICRC07Roth} is exploited to obtain the
zenith angle correction: it assumes that the cosmic ray flux is isotropic in 
local coordinates, i.e. the number of events above a certain threshold energy
is constant as a function of $\cos^2 \theta$. 
\begin{figure}
\begin{center}
\psfig{figure=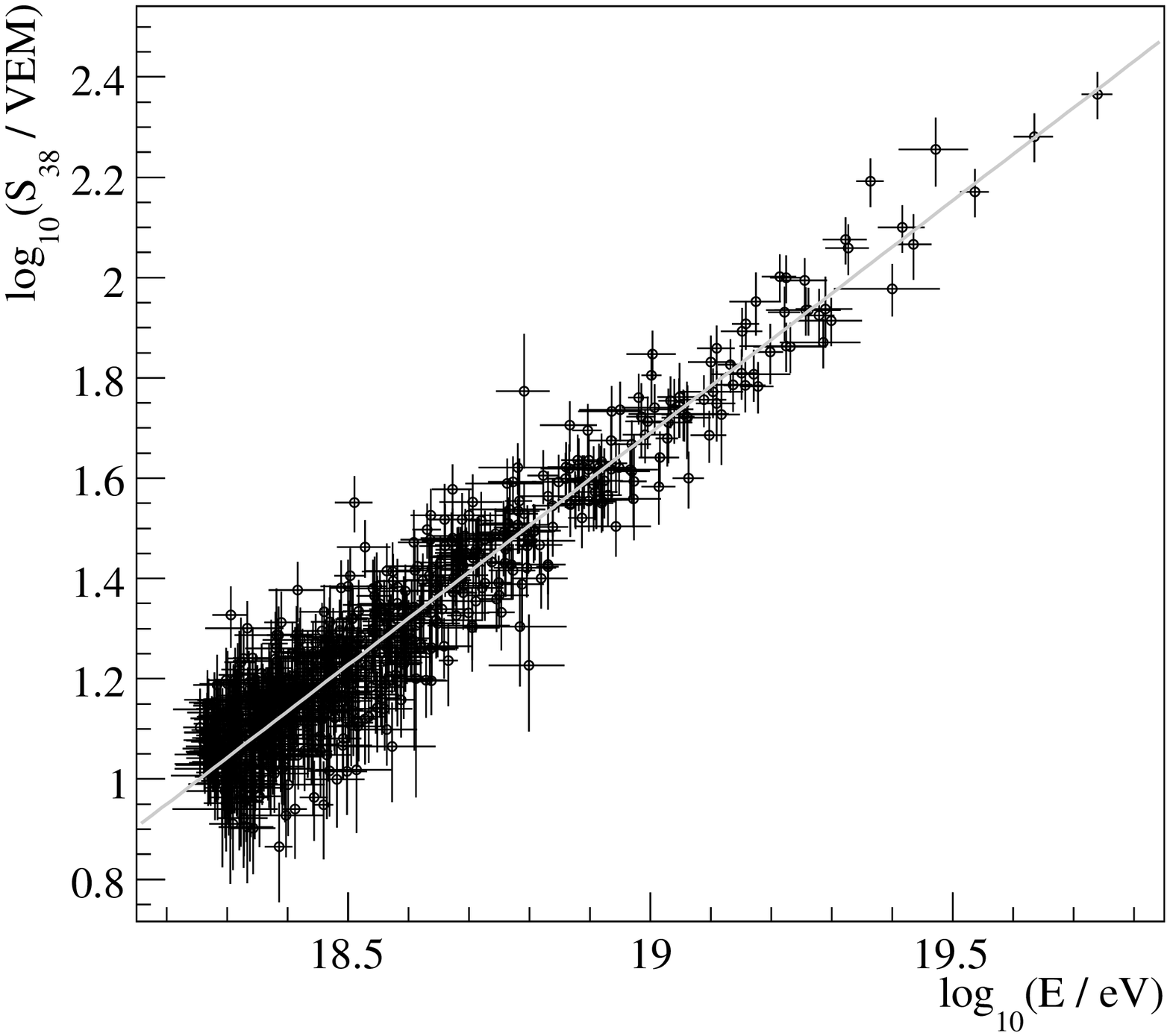,width=0.49\textwidth}
\psfig{figure=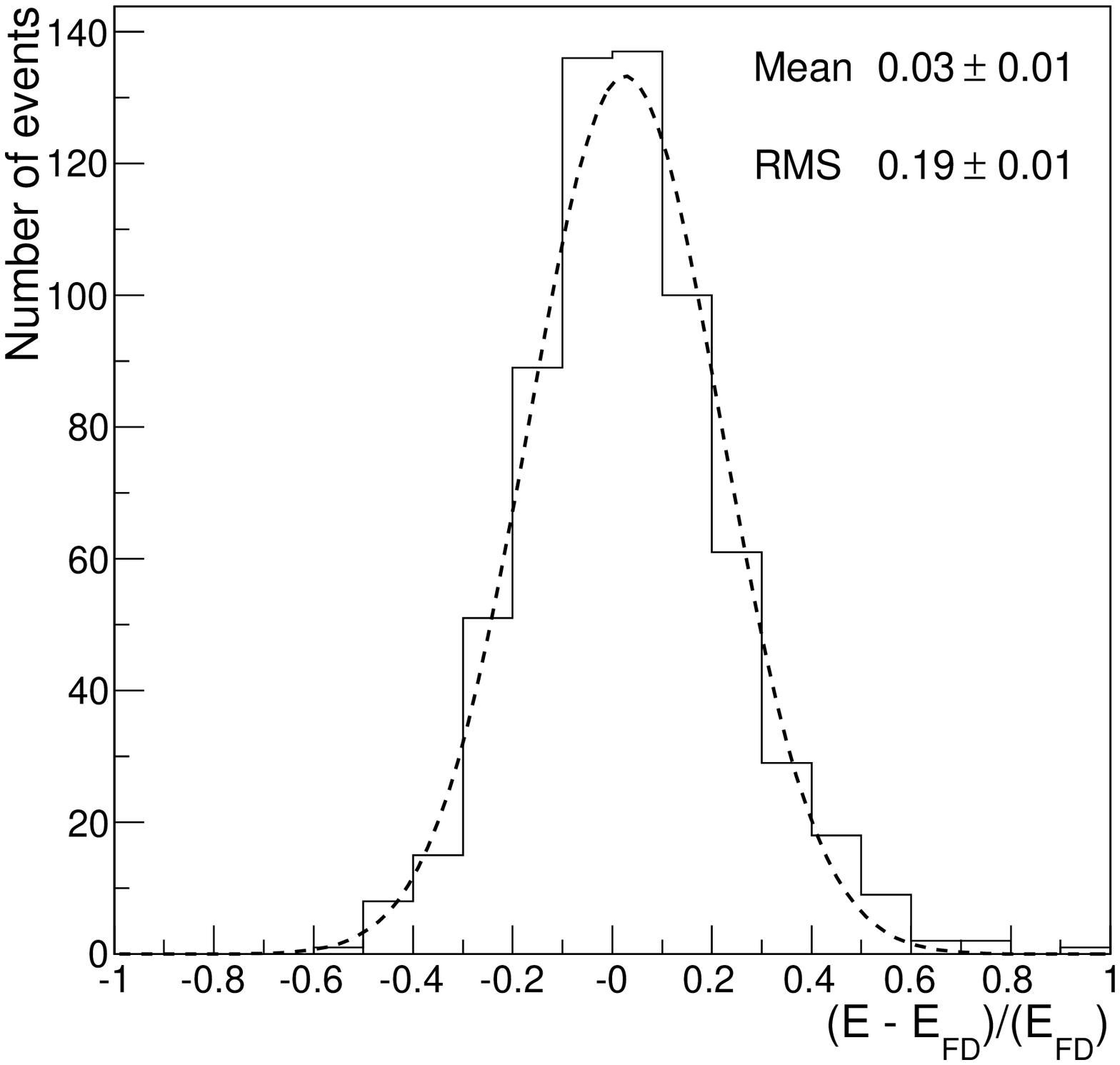,width=0.49\textwidth}
\end{center}
\caption{(left hand side) Energy
  calibration.  The relation between $S_{38}$ and
  energy is almost linear and is shown with the continuous
  line.
  (right hand side) The fractional difference between the assigned energy and
  the  FD  energy.~\protect\cite{bib:ICRC07Roth} \hfill
\label{fig:en_calib}}

\end{figure}
This hypothesis leads to  the correction
function for S(1000) shown in  Fig.~\ref{fig:cic}(left). It is a
second degree polynomial $S(1000)(x)= S_{38}\cdot(1+a\cdot x+ b\cdot
x)$ where $x=\cos^2 \theta- \cos^2 38^\circ$,  $a=0.94\pm 0.06 $ and
$b=-1.21\pm0.27$. The quantity  obtained by using the empirical fit
shown in the same 
figure, $S_{38}$, represents the  signal at \unit[1000]{m} the very
same  shower would have produced if it had arrived from a  zenith
angle of $38^\circ$. This angle corresponds to the median of the
zenith angle distribution of the SD data. The number of events above a
certain $S_{38}$ is zenith angle independent. In principle the
attenuation might  be  energy dependent, because showers with higher
energies develop deeper in the atmosphere and can be observed before
their maximum development. This effect was found to be negligible.
Distributions of the number of events for equal $\cos^2\theta$ intervals as
a function of S(1000) and $S_{38}$ are shown in
Fig.~\ref{fig:cic}~(right). After applying the correction function the
zenith angle dependence disappears.  

The transformation from $S_{38}$  to energy is obtained by so called
golden hybrid events. These are air showers that triggered both
the SD array and the FD. 
%The 
%energy reconstruction from the longitudinal profile is a complex
%process, mainly due to unawareness of the exact atmosphere conditions,  
%therefore only high quality data are used for an energy
%assignment. 
Only a subsample of high quality FD measurements is used for the 
energy calibrations: For example the reduced $\chi^2$ for the fit is required
to be less than 2.5, data are selected only if measurements of the 
vertical aerosol optical depth are available and the maximum of the
shower development is required to be in the field of view of the
detector. Moreover, only events with a fraction of Cerenkov contribution of
less than 50\% of the total light, an uncertainty on the position
of the shower maximum smaller than \unit[40]{g/cm$^{2}$} and a
relative total energy  uncertainty less than 20\% are used in the calibration
procedure.

The relation between $S_{38}$ and the FD energy is shown in
Fig.~\ref{fig:en_calib}~(left). As can be seen, it
 exhibits a power law  correlation with a
relative dispersion of $19\pm 1\%$ (Fig.~\ref{fig:en_calib}~(right)).
The uncertainties in the determination of both FD energy and SD
signal are assigned on an event by event basis. The energy uncertainty
($\approx$ 8\%) 
includes also propagated  atmosphere uncertainties and uncertainties
from the air shower geometry reconstruction. 
The $S_{38}$ uncertainty($\approx$ 16\%) contains the lateral distribution
function  
assumption, the shower to shower fluctuation and reconstruction
accuracies. The best fit gives the transformation from $S_{38}$ to
energy as $E= A \cdot S_{38}^{B}$, $A= 1.49\pm
0.06({\rm stat})\pm 0.12 ({\rm sys})[10^{17} eV]$ and $B= (1.08\pm 0.01({\rm
  stat})\pm (0.04)({\rm sys}))$ with a reduced $\chi^2$ of 1.1.

\begin{figure}
\begin{center}
\psfig{figure=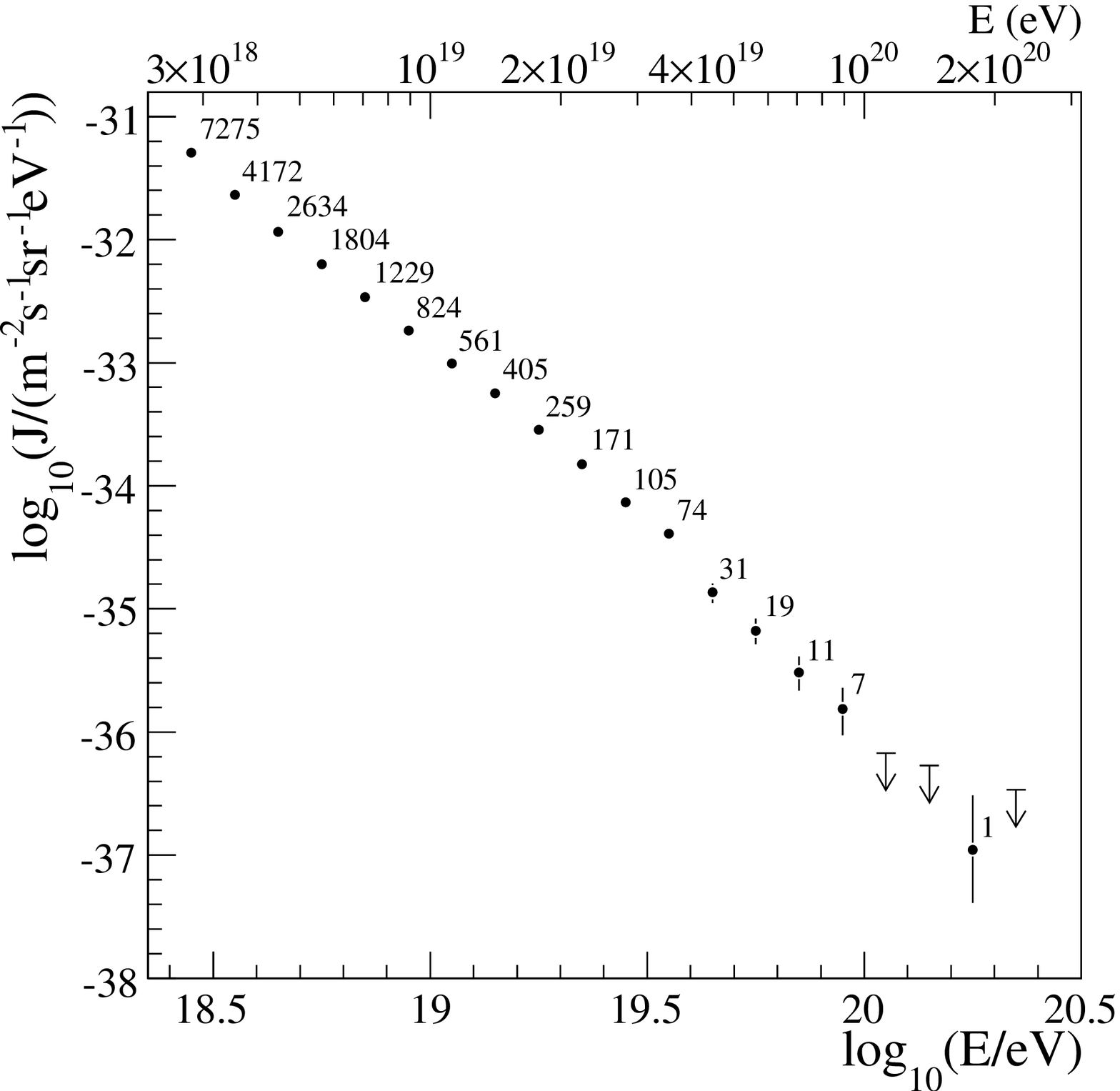,width=0.49\textwidth}
\psfig{figure=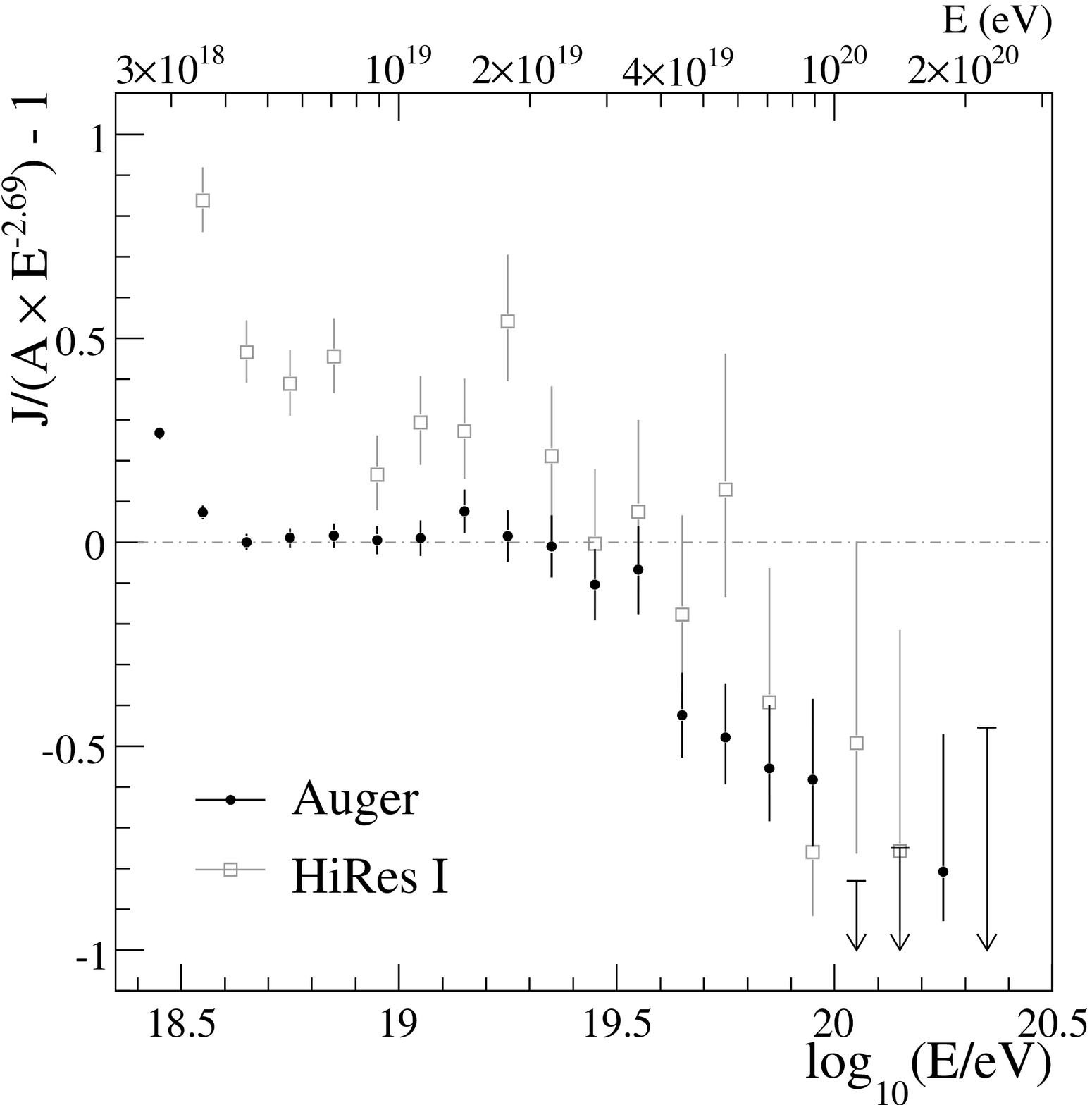,width=0.49\textwidth}
\caption{(left hand side) The differential flux as a function of
  energy. Also shown are the number of events in each bin. (right hand
  side) The fractional  difference between the Auger vertical and
  HiRes I data and a    spectrum with an index of
  2.69~\protect\cite{bib:ICRC07Roth}. \hfill   
\label{fig:sdspectr}}
\end{center}
\end{figure}

The spectrum built from 
%% MU really only twenty thousand events??
$2\cdot 10^4$ 
events recorded until August 2007 is
shown in Fig.~\ref{fig:sdspectr}. The acceptance is computed by
simple geometrical considerations and from the continuous monitoring
of the configuration of the array~\cite{{bib:ICRC05AllardTrigger}}. The data 
set used for obtaining the energy spectrum contains only events with energies
greater than \unit[$3\cdot 10^{18}$]{eV}, since only above this 
energy the array is fully
efficient.    
Due to reconstruction and trigger efficiency issues only events with a
zenith angle of less than $60^\circ$ are included in analysis. The
integrated exposure for this period is \unit[$7\cdot 10^3$]{km$^2$ sr yr}. The
uncertainty of the acceptance is less than 5\%. Having a  duty cycle
of almost 100~\% the vertical  spectrum from the SD has the lowest
statistical and systematic uncertainties.

In Fig.~\ref{fig:sdspectr}~(right)  the fractional difference
between the vertical spectrum and a power-law  $\propto E^{-2.69}$ is
illustrated.    
Two spectral features are clearly visible: the so-called {\it ankle} at
energies of \unit[$\approx 10^{18.5}$]{eV} and a flux suppression at energies
above  \unit[$\approx 10^{19.6}$]{eV}.
A continuation of the  spectrum as a power law with   index
$2.69$  at highest energies predicts $167\pm3$ events above
\unit[$10^{19.6}$]{eV} and 
$35\pm 1$ above \unit[$10^{20}$]{eV}, whereas we observe only 69
events and 1 event. The hypothesis of a pure power-law can be rejected
with a significance of $6$ standard deviations, independent of the
energy scale uncertainties.

\section{The Auger spectrum: combining the hybrid and surface detector
measurements}

\begin{figure}
  \begin{center}
    \begin{minipage}{0.49\textwidth}
      \vspace*{0.75cm}
      \psfig{figure=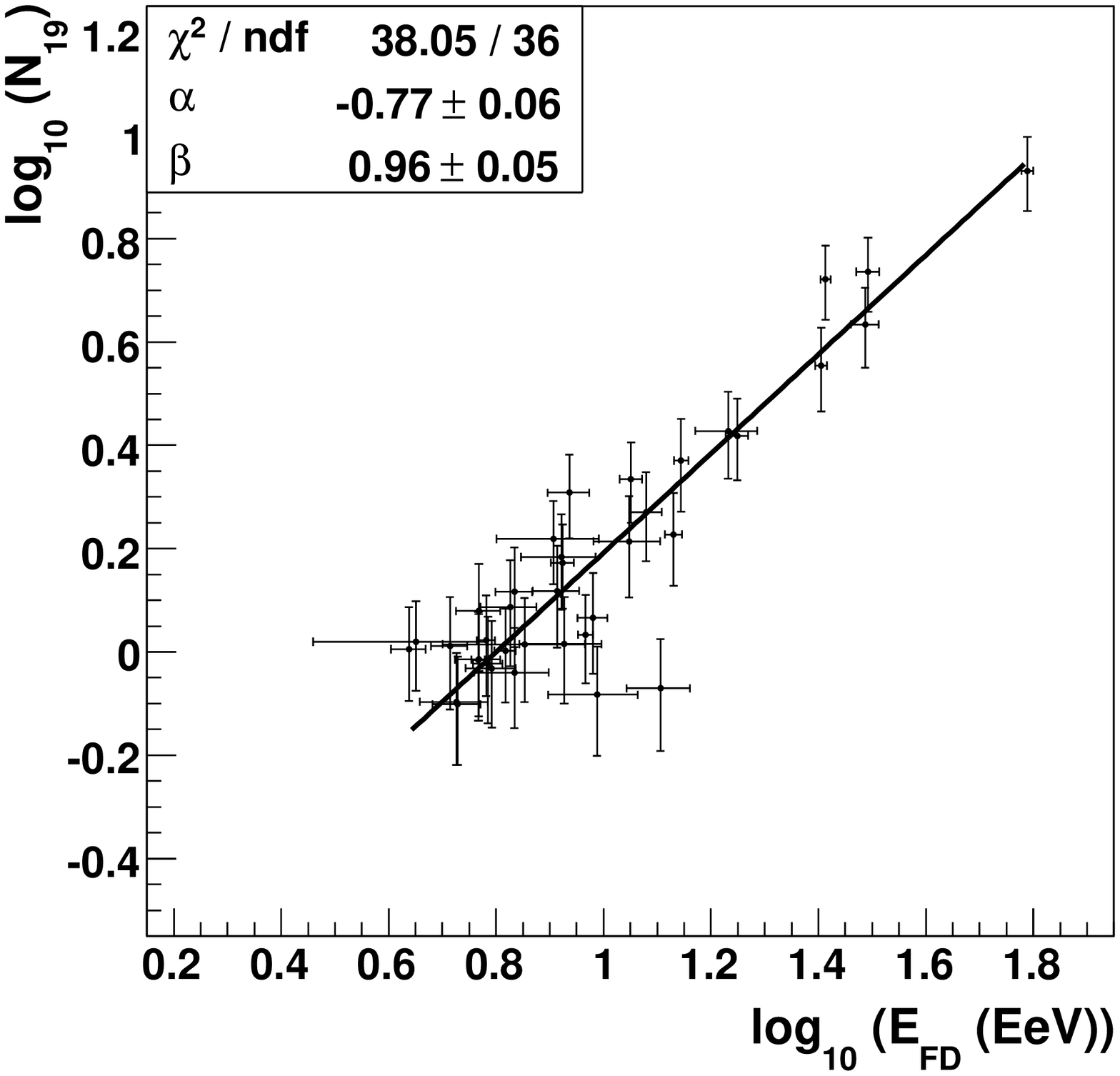,width=\textwidth}
    \end{minipage}
    \begin{minipage}{0.49\textwidth}
      \psfig{figure=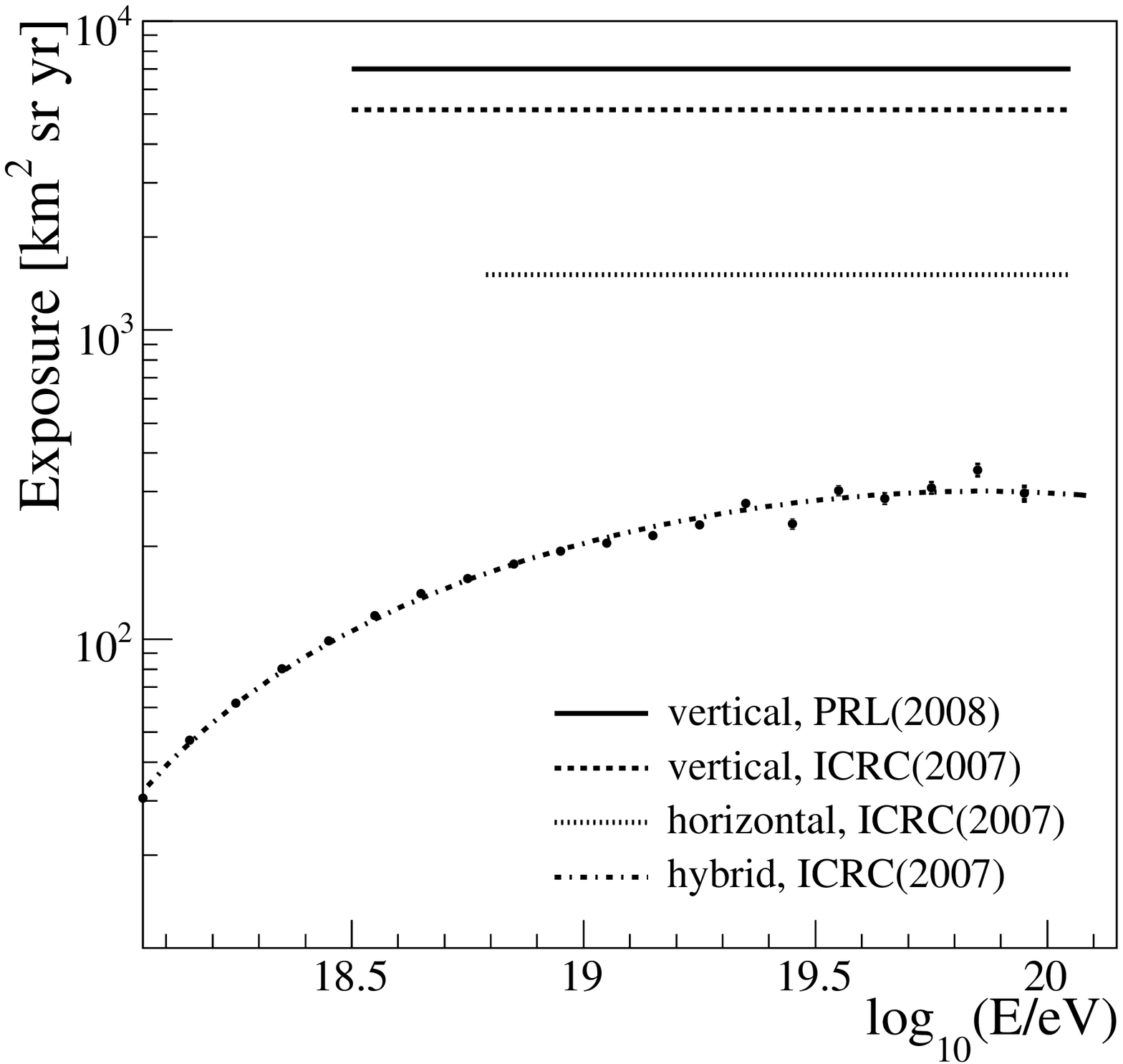,width=\textwidth}
    \end{minipage}
    \caption{(left hand side) The energy calibration for the surface
    detector events with zenith angle between $60^\circ$ and
    $80^\circ$. (right hand side) The exposure for the three data sets
    used to determine the cosmic ray flux. \hfill
      \label{fig:hybandexp}}
  \end{center}
\end{figure}

Air showers measured by the SD array with a zenith
angle between  $60^\circ$ and $80^\circ$ are  used to determine an
independent spectrum. The procedure to derive the energy
is equivalent to the vertical events, but instead of using $S_{38}$ the shower
size is determined from the relative distributions of the two-dimensional
muon number densities at ground level~\cite{Facal San Luis:2007it}.
The normalization factor of the muon map, $N_{19}$, gives the total
number of muons relative to a shower initiated by a proton with an
energy of \unit[$10^{19}$]{eV}. The relation between $N_{19}$ and the
hybrid energy is shown in Fig.~\ref{fig:hybandexp}. The statistics is
rather low compared to the vertical energy calibration, but a clear
almost linear dependency is seen.

The acceptance calculation for this set of events is purely geometrical and 
the threshold energy above which the trigger efficiency is more than 98\% is
\unit[$6.3\cdot 10^{18}$]{eV}. Above this energy the integrated exposure until
the end of February 2007 is  \unit[1510]{km$^2$ sr yr}; 29\% of the equivalent
acceptance for vertical events~\cite{Facal San Luis:2007it}.

%% The data collected at the Pierre Auger are divided in three sets. The first
%% set consists of  air showers with zenith angle of less than
%% $60^\circ$  detected by SD. The energy calibration described above is applied
%% in this case. The integrated 
%% exposure  reported here is \unit[5165]{km$^2$ sr yr} after some quality cuts
%% (through February 2007), more than a factor of three larger than the exposure
%% obtained by the largest forerunner experiment
%% AGASA~\cite{bib:AGASA}. (Elsewhere in these Proceedings, Bonino reports other
%% Auger results using an exposure through August 2007 of \unit[9000]{km$^2$ sr
%%   yr}).   

Another data set available to measure the cosmic ray flux contains
hybrid events. 
The hybrid exposure calculation relies on
an accurate simulation of the fluorescence detector and the atmosphere.
% and single tank response and 
% it is energy dependent. 
A large sample of Monte Carlo simulations are performed  to reproduce
the exact  conditions of the experiment and the entire sequence of
given configurations, from camera pixels to the combined SD-FD data
taking of the observatory. 
The rapidly growing array, as well  as the seasonal and  instrumental
effects are reproduced in the simulations within \unit[10]{min} time
intervals. 
The systematic uncertainty in the hybrid spectrum is currently dominated 
by the calculation of the exposure and reaches 20\% in the low energy
range.
The advantage of the hybrid measurement of the energy
spectrum~\cite{Perrone:2007he} is the coverage of the energy range between
\unit[$10^{18}$]{eV} and \unit[$3\cdot 10^{18}$]{eV}. 

The exposure for the three data sets is illustrated in
Fig.~\ref{fig:hybandexp}. The hybrid measurement extends to the lowest
energy range. Including the use of inclined SD data improve the
statistics noticeably  as this set has about a quarter of the event
totals of the vertical data set in the highest energy region. 
%% The inclined SD  data increase with almost one third the
%% event statistics of the vertical ones in the highest energy region. 

All spectra are affected by the 22\% uncertainty in the FD energy 
scale, the main contributions coming from the determination of the
fluorescence yield (14\%). 
The profile reconstruction itself  contributes with 10\%.
The absolute calibration of the telescopes is done every few months and
 has a contribution of 9.5\%. The correction from the vertical aerosol optical
 depth measurements is 5-18\%, giving an uncertainty on the energy of
4\%.  The advantage of the 
hybrid energy determination over the only SD assignment (e.g. as in the
case of AGASA experiment) is that the uncertainties are experimentally
driven and can be improved. A single surface detector experiment determines
the primary cosmic ray energy with the help of simulations of air
showers therefore the uncertainties are driven by theoretical
uncertainties which are harder to decrease. 

\begin{figure}
\begin{center}
\psfig{figure=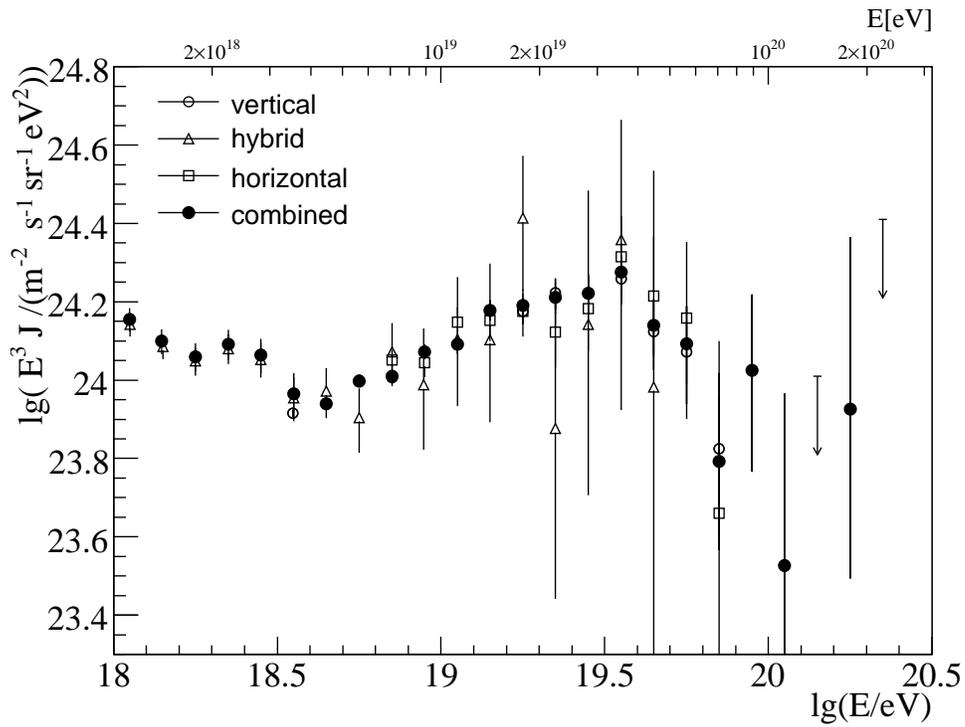,width=0.85\textwidth}
\caption{The energy spectra obtained at the Pierre Auger
  Observatory as presented in \protect\cite{Yamamoto:2007xj}. The combined spectrum is shown with filled
  circles. Open circles represent the energy spectrum obtained from
  events with a zenith angle of less than $60^\circ$, triangles
  represent the hybrid measurement and open squares the energy
  spectrum obtained from surface detector data with a zenith angle
  between $60^\circ$ and $80^\circ$. Only statistical uncertainties
  are shown. \hfill
\label{fig:allspectr}}
\end{center}
\end{figure}

The energy spectra obtained with the three methods are illustrated in
Fig.~\ref{fig:allspectr}. The systematic
uncertainty of the energy does not affect the relative comparison of the three 
spectra.  The agreement is well within the independent 
systematic uncertainties, the difference between the overall
normalizations is at a level of less than  4\%.  

The Auger combined differential flux, shown in the
same figure multiplied with the third power of energy,  is extending
over the widest energy range possible and with 
minimal uncertainties. To deduce the spectrum a maximum
likelihood method is applied taking into account the  independent
uncertainties of each measurement.~\cite{Yamamoto:2007xj} The combined
spectrum is dominated by the vertical surface detector measurement
above \unit[3]{EeV} and by the hybrid spectrum in the lower energy
range.  The spectral index changes from $\gamma_1=-3.30\pm0.06$ to
$\gamma_2=-2.69\pm0.02({\rm stat})\pm 0.06 ({\rm   sys})$ at
$\log(E_{\rm   ankle}/{\rm eV})=18.65\pm 0.04$, and above
\unit[$10^{19.6}$]{eV} to $\gamma_3=-4.2\pm0.4({\rm stat})$.

\section{Conclusions}\label{sec:conc}

The energy spectrum has been measured at the Pierre Auger Observatory
with three independent data sets and the agreement is better than
4\%. 

A flux suppression at the highest energies has been established with a
significance of 6 standard deviations.  Combined with the recent
result that the highest energy cosmic rays are anisotropic and a good
correlation only with relatively nearby sources has been
found~\cite{bib:Science}, it hints towards a GZK effect. 

A change of the spectral index occurs at about  \unit[4]{EeV}
which might indicate either the transition from galactic to
extragalactic origin of cosmic rays or a propagation effect. To
distinguish between the models a determination of the mass
of the cosmic rays~\cite{bib:ICRC07Unger}  over the entire energy
range is necessary.  
The exact location of the spectral index change and an accurate measurement of
the spectrum below \unit[3]{EeV} will be possible in the near
future with AMIGA and HEAT extensions of the observatory. 

The exact shape of the energy spectrum in the highest energy range,
which can be used to distinguish between acceleration
scenarios~\cite{Yamamoto:2007xj}, will be determined with greatly
improved precision over the next 10 years of data taking. 

%% One year of complete Observatory data will
%% deliver the same statistics as presented here.

%%  If in the planned
%% northern side Observatory  different spectrum indices will be
%% measured, a strong hint to the acceleration sites can be given as the
%% matter distribution in the northern sky differs from the southern part.  

\end{document}